\documentclass[pre,twocolumn,preprintnumbers,showpacs,floatfix,superscriptaddress]{revtex4-1}
\usepackage{epsf} 
\usepackage{graphicx}% Include figure files
\usepackage{dcolumn}% Align table columns on decimal point
\usepackage{bm}% bold math
\usepackage{subfigure}
\usepackage{color}
\usepackage{amsthm}
\usepackage{amssymb}
\usepackage{amsmath}
\usepackage{pb-diagram}
\usepackage{epstopdf}

\usepackage{hyperref}

\definecolor{gray85}{gray}{0.85} % 15%
\definecolor{gray8}{gray}{0.8} % 20%
\definecolor{gray7}{gray}{0.7} % 30%
\definecolor{gray6}{gray}{0.6} % 40%
\definecolor{gray5}{gray}{0.5} % 50%
\definecolor{gray4}{gray}{0.4} % 60%
\definecolor{gray35}{gray}{0.35} % 65%

\begin{document}

\preprint{}

\title[]{Characterizing Granular Networks Using Topological Metrics}

\author{Joshua A. Dijksman}
\affiliation{Physical Chemistry and Soft Matter, Wageningen University, Wageningen, The Netherlands}

\author{Lenka Kovalcinova*}
\affiliation{Department of Mathematical Sciences, Center for Applied Mathematics and Statistics, New Jersey Institute of Technology, Newark, NJ 07102, USA}

\author{Jie Ren}
\affiliation{Merck Research Laboratories, Merck \& Co., Inc., West Point, PA 19486, USA}

\author{Robert P. Behringer}
\affiliation{Dept. of Physics, Duke University, Science Drive, Durham NC 27708-0305, USA}

\author{Miroslav Kramar}
\affiliation{Advanced Institute for Materials Research, Tohoku University, Sendai, Japan},

\author{ Konstantin Mischaikow}
\affiliation{Department of Mathematics, Rutgers University, Piscataway, NJ 08854-8019, USA},

\author{Lou Kondic}
\affiliation{Department of Mathematical Sciences, Center for Applied Mathematics and Statistics, New Jersey Institute of Technology, Newark, NJ 07102, USA}


\begin{abstract}
We carry out a direct comparison of experimental and numerical realizations of the exact same granular system as it undergoes shear jamming. We adjust the numerical methods used to optimally represent the experimental settings and outcomes up to microscopic contact force dynamics. Measures presented here range form microscopic, through mesoscopic to system-wide characteristics of the system. Topological properties of the mesoscopic force networks provide a key link between micro and macro scales. We report two main findings: the number of particles in the packing that have at least two contacts is a good predictor for the mechanical state of the system, regardless of strain history and packing density. All measures explored in both experiments and numerics, including stress tensor derived measures and contact numbers depend in a universal manner on the fraction of non-rattler particles, $f_{NR}$. The force network topology also tends to show this universality, yet the shape of the master curve depends much more on the details of the numerical simulations. In particular we show that adding force noise to the numerical data set can significantly alter the topological features in the data. We conclude that both $f_{NR}$ and topological metrics are useful measures to consider when quantifying the state of a granular system.
  \end{abstract}

\maketitle

An important class of particulate systems includes granular materials, colloids, foams and molecular glass formers. These materials can become ``rigid'', or ``jammed'', in the absence of long range spatial order. Jammed states are solids in mechanical equilibrium, with non-zero elastic moduli~\cite{liu98, vanhecke}. In order to be in mechanical equilibrium, forces must be transferred from particle to particle, creating self-organized networks of contacts and forces.
These networks undergo rearrangements  when shear induced deformation of the materials occurs.
The networks are typically spatially heterogeneous~\cite{camboubook}, and during deformation, they are temporally intermittent, with avalanches, slip events, fracture and elasto-plastic failure modes. Finding the microscopic metrics responsible for this broad spectrum of mechanical behaviors across the solid-liquid transition has been challenging, since many of the conventional tools for characterizing ordered thermal systems do not apply.

\begin{figure}[!t]
\includegraphics[width=8.0cm]{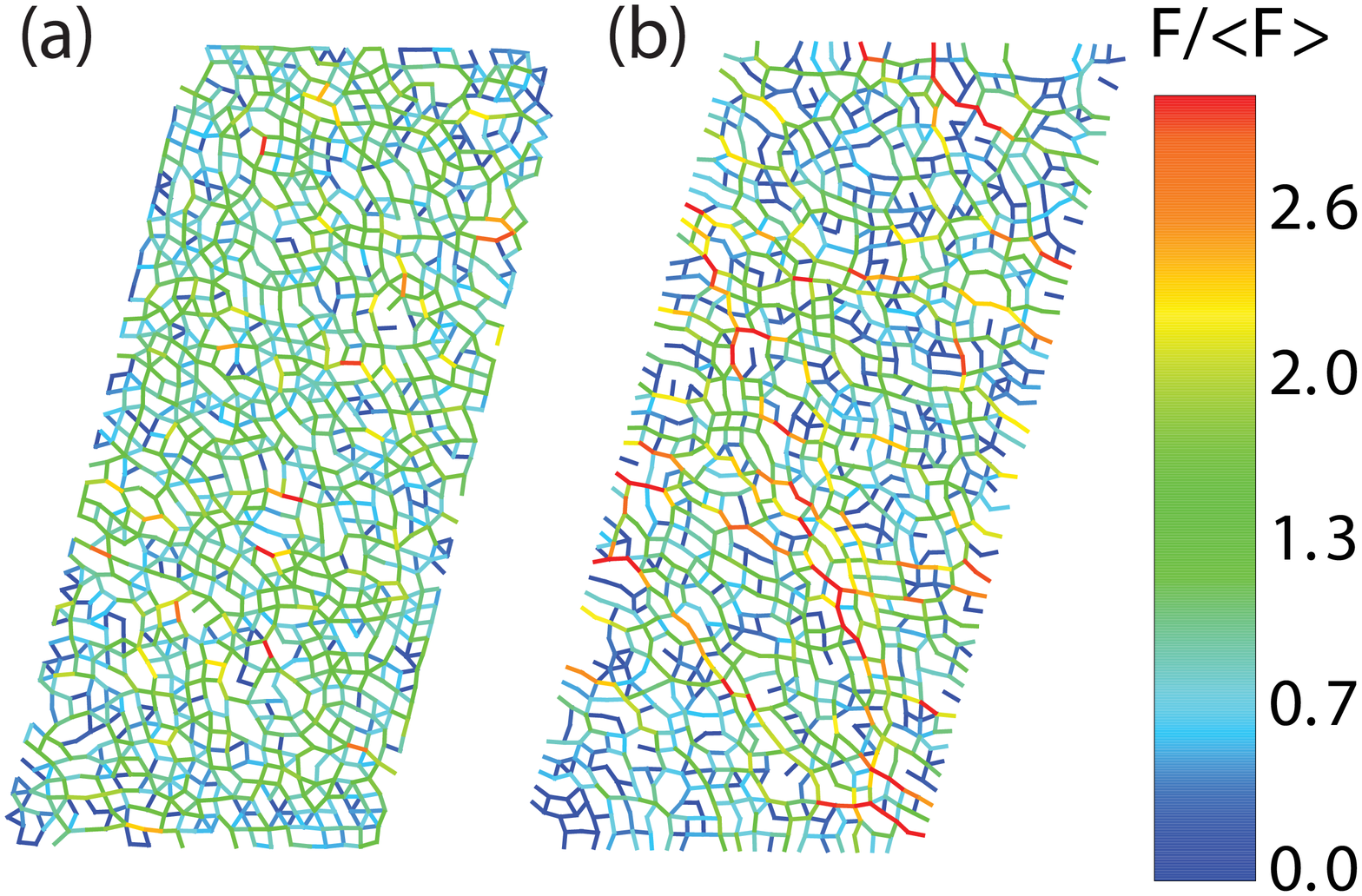}
\caption{\label{fig:0} Examples of an experimental (a) and a numerical (b)
  force network. Experimental and numerical examples are generated at the same strain
  amplitude $\gamma = 14.9\%$ and $\phi = 0.8036$. The color scale represents the
  total force at each contact, normalized by the concurrent mean
  force. As described in the text, the simulation results in
  this and all the following figures include additional noise of amplitude $0.01$ N,
  except if specified differently.
    }
\end{figure}

Understanding how granular materials self-organize into mechanically stable states requires consideration of the structure of force networks.  Figure~\ref{fig:0}
shows examples of typical force networks found in the experiments and simulations. A full description of the force network requires a high dimensional space that reports on microscopic features and does not directly reveal its mesoscopic structure. In order to understand this structure, we need statistical tools for their characterization that are sensitive, systematic, unbiased, minimal, and consistent with macroscopic properties, such as the system-wide stresses. These tools must also be able to distinguish different states of the system, based on force networks.

To demonstrate the need for mesoscopic descriptions of the force network we will consider the granular response to quasi-static shear strain, $\gamma$, which is analogous to time in a more conventional dynamical system. The structure of the network as the granular system is strained shows a sensitive dependence on the initial conditions. We show that simple topological measures detect the variability in force networks formed in response to shear in quasi-two-dimensional granular systems. These metrics capture features that are missed by conventional measures, such as stresses or contact force distributions.

In this article, we consider two replicas of the same granular system: experimental and computational. The studies of these realizations show the importance of the fraction of non-rattler particles, $f_{NR}$, {\bf defined here for frictional particles,} as particles with at least two contacts,  as a relevant state variable. In particular, the considered metrics fall onto master curves when expressed as functions of $f_{NR}$.
These metrics include traditional measures, such as contact numbers, stresses, and contact force probability distribution functions which identify system-wide as well as microscopic features. We observe a good match between the experimental and  computational systems at these scales. To our knowledge, this is a first attempt to directly compare the microscopic and network properties of experiments and simulations on, nominally, the same system. However, Veje et al.~\cite{veje98} made comparisons of macroscopic properties for experiments and simulations of a Couette system.

However, at the mesoscale, there are clear differences identified using topological metrics.
We conjecture that the differences arise from small imperfections and noise in the experiments that are absent in the simulations.  The topological metrics that we  employ are sensitive to the force noise in the experimental data sets and capture these differences.  Our conjecture is reinforced by the fact that by introducing appropriate random noise in the
simulations we can generate a reasonable match between the statistics of the topological networks in simulations and experiments.   This finding suggests that the considered topological
measures may have an important utility in quantifying the properties of
intrinsic noise that is always involved in the experiments.
As a consequence, topological techniques provide new opportunities for identifying the scale of the experimental noise, and distinguishing noise from intrinsic fluctuations.

\section{Topological metrics} Although there exists a number of tools to extract network information, see~\cite{barabasi,alexander98} for reviews, recent work~\cite{peters05,tordesillas_pre10,carlsson12,kondicepl12,arevalo13,arevalo14,kramar2013,daniels_pre12} demonstrates that topological methods are well suited to characterize force networks in granular packings.  In this work, we use the Betti numbers $\beta_0$ and $\beta_1$ to quantify the topology of a network, where $\beta_0$ indicates the number of clusters or connected components of the network, and $\beta_1$ characterizes the number of loops. Thus, in a rough sense, $\beta_0$ is a measure of force network segments, and $\beta_1$  is related to their interconnectivity. Computations were performed using software Perseus~\cite{perseus}. In particular, we are interested in the properties of the (simplicial complex) network that describes the force interactions between the particles with force magnitude larger than $f_c$. We pick $f_c = 1.0\langle |\textbf{f}| \rangle$ in our
analysis (here $\langle |\textbf{f}| \rangle$ is the average force), and note that results are not very sensitive to this choice of threshold. This network is constructed in the following manner. Every particle, $p_i$, is represented by a vertex, $v_i$. The edge $<v_i,v_j>$  belongs to the network  if  the magnitude of the force interaction, between the corresponding particles $p_i$ and $p_j$, is larger than $f_c$. In our construction, we follow the similar approach as in previous works, see e.g.~\cite{kramar2013,physicaD14} and ignore the ``trivial’' loops forming between three particles in contact. Therefore the boundary of the loops detected by $\beta_1$ must contain  at least four edges.

While more elaborate tools from algebraic topology have been used to quantify force networks in simulations~\cite{poly1,poly2}, in the present work we restrict ourselves to the simple measures described above. The reasons for this choice are twofold: \emph{(i)} This method allows for a direct comparison of the  statistical properties of force networks between simulations and experiments.  To our knowledge, such comparison has not been attempted so far. \emph{(ii)} These methods show potential for assessing the type and size of the noise present in the data~\cite{Omer3}.

\begin{figure}[!t]
\includegraphics[width=8.5cm]{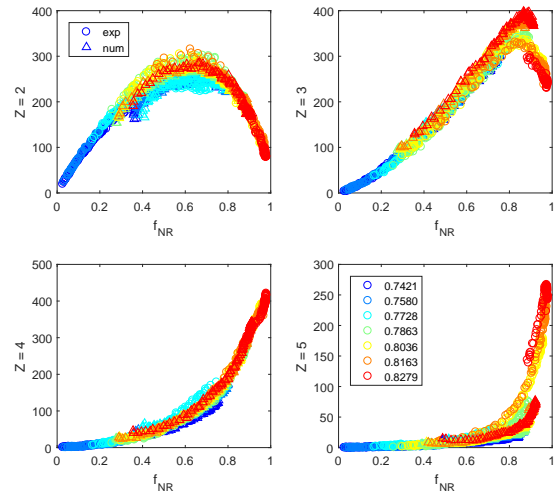}
\caption{\label{fig:1} The evolution of the fraction of particles with
  contact number $Z_{2,3,4,5}$ as a function of the fraction of
  non-rattlers $f_{NR}$, for both experiments ($\circ$) and simulations
  ($\bigtriangleup$) for a range of $\phi$'s (color
  scale).
 }
\end{figure}

\section{Experimental \& Numerical settings} The experiments involve
shear of constant-density two dimensional packings of $\sim 1000$
photoelastic bidisperse disks, starting from a stress free state. The
packing density is given in terms of the packing fraction, $\phi$. The
particles interact via force laws that include friction, and the
result is that the applied strain shear-jams the
packing for lower $\phi$ than the isotropic jamming value~\cite{bi11}. This protocol allows one to probe a large range
of mechanically distinct states, from very loose to highly jammed
packings. In particular, we use experimental data obtained in a simple
shear geometry described in \cite{ren13, ren11} -- see the Appendix for details.  The particles rest on
a co-moving, articulated base, that through its boundaries induces a
linear shear profile, suppressing shear bands and other large scale
inhomogeneities. We consider results for packings in the range $0.77 \leq \phi \leq 0.825$, where $\phi = 0.77$ is the lowest packing fraction for which we achieve shear jammed states with this apparatus.
We then quasi-statically shear the system by a
sequence of $100$ strain steps of 0.27\% each, for a maximum
strain of $\gamma = 27\%$~\cite{ren11}, and extract
the information about all the forces the particles experience at each strain step~\cite{ren13}. The force extraction algorithm is applied to each particle individually. Hence, every contact between two particles yields two force vectors. In the topological analysis we use the average of the norms.
Note that  this approach has its limitations. For larger pressures, $P$,  we reach a force level such that
deformations at contacts become large.  We note that the force inverse algorithm assumes small deformations at contacts to make the process more efficient and
becomes increasingly inaccurate when $f_{NR} > 0.95$ where deformations are large, so we omit this (small) data set from the analysis where forces are involved (see Appendix for details). The exact value of this cutoff is not crucial for the present purposes. Note that this is not an intrinsic limitation on the technique; however,  since we are mostly interested in behavior near jamming where forces are moderate, amending current algorithms is not necessary.

{\em Simulation Details} Despite the simplicity of the numerical scheme, it is highly nontrivial to select the right packing preparation protocol. This sensitivity of granular mechanics to initial preparation is well known even for frictionless systems \cite{goodrich2014} yet makes a direct comparison between numerical and experimental results very nontrivial. Note that regardless of the preparation protocol chosen, the qualitative behavior observed was always similar; the collapse of all data sets as a function of $f_{NR}$ is particularly robust. Two features however required protocol fine tuning: the density \emph{range} and \emph{offset} in which the shear jamming and network anisotropy were observed and the network anisotropy. The fine tuning allowed us to match the both range and offset between experiments and numerics. The contract number and pressure dynamics are found not to be influenced by preparation protocol.

The simulations use a soft-particle non-linear force model and
reproduce the experimental settings as closely as possible. The
shear geometry, particle sizes and numbers, friction coefficients,
density and elastic moduli used in the numerical simulations match the
experimental values as taken from the current data or previous work~\cite{ren13,bi11}. The inelasticity of the particles is modeled
through a coefficient of restitution and frictional interaction between
the particles using the Cundall-Strack model~\cite{DEM_ori}. We also
simulate a moving base, including translational and rotational
friction between the base and the particles.  The
details of implementation can be found in the Appendix. The simulations reported here
are carried out for the same values as for which we have experimental data. We show
below that simulations produce results that are comparable to experiments.

As we will discuss in more detail later, we add random
noise to simulation data: all the figures in the paper include this additional
noise, unless otherwise specified.   Importantly, the essential results shown in
Figs.~\ref{fig:0}~-~\ref{fig:23} are insensitive to the noise addition.

\section{Common features of experiments and simulations} We extract
particle positions and inter-particle forces for each particle at every 0.27\% strain
step, and we compute the local stress tensor
$\underline{\sigma}$ using the Irving-Kirkwood
method~\cite{irving-kirkwood}. This provides the pressure from the sum
of its eigenvalues $P = (\sigma_1 + \sigma_2)/2$ the shear stress,
$\tau = (\sigma_2 - \sigma_1)/2$ and the stress anisotropy,
$\tau/P$. At each $\phi$, we carry out five realizations in both simulations and experiments, and average the results.

To illustrate the degree of agreement between experiments and
simulations when conventional measures are considered, we start by
exploring the average number of contacts per particle, $Z$. This is an important
quantity since there is a minimum or isostatic value of $Z$, $Z_{iso}$ for
marginal stability. For frictional particles, $Z_{iso} = N + 1$, (modulo a small correction due to the constraint imposed by Coulomb friction), where
$N$ is the system dimension, e.g. $Z_{iso} = 3$ for 2D frictional
disks.  If a system is sheared from zero stress into a shear jammed
state, $Z$ must reach at least $Z_{iso} = 3$ when the system becomes jammed.  However, the parts of the force network
that first form during the shear jamming process contain a number of
particles with only two contacts. Hence, it is relevant to
consider not only $Z$, but also $Z_n(\phi,\gamma)$, the fraction of
particles with $n$ contacts.

A key finding in \cite{bi11} is that many properties such as stresses and $Z$ depend on  $f_{NR}$,  in a universal way.  Here we show that $f_{NR}$ also determines the dynamics of $Z_n(\phi,\gamma)$. Figure~\ref{fig:1} shows data for $Z_{2,3,4,5}$ vs. $f_{NR}$.  There are two outstanding features in this data: \emph{(i)} the dynamics of each $Z_n$ collapses on a single curve, independent of $\phi$; and\emph{(ii)} the agreement between experiments and simulations is quantitative.  We thus conclude that the simulations reproduce the experiments very well, and that $f_{NR}$ can be used as an apparently universal state variable to describe the state of the system.

\begin{figure}[!t]
\includegraphics[width=8.5cm]{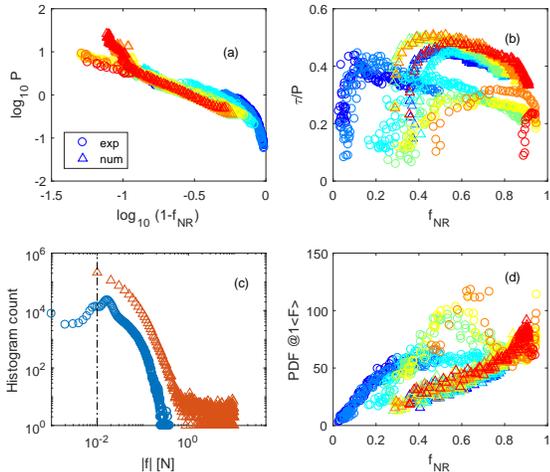}
\caption{\label{fig:23} (a) Pressure evolution versus 1-$f_{NR}$ for
  both experiments ($\circ$) and simulations ($\triangle$).   (b) Normalized anisotropy $\tau/P$
  evolution as a function of $f_{NR}$. (c) histogram of contact force magnitudes $|f|$ for all contacts in the five runs at $\phi = 0.7863$ for both numerics and experiments. The dot-dashed line indicates the force noise added after the numerical runs were completed. (d) The probability
  distribution function of the norm of the contact forces
  $P(|\textbf{f}|)$ as a function of $f_{NR}$ for both experiments
  ($\circ$) and simulations ($\triangle$) for the force bin centered
  around $1\langle|\textbf{f}| \rangle$.}
\end{figure}

Figure~\ref{fig:23} shows other conventional measures comparing the experimental and the numerical data: in
Fig.~\ref{fig:23}a we see that the experimental pressure data for all $\phi$'s
collapse to a single curve.  Furthermore, we find power
law scaling for $P$ vs. $1-f_{NR}$.  Intuitively, the inverse relation
between $P$ and $f_{NR}$ makes qualitative sense: the larger the
fraction of rattlers, $1-f_{NR}$, the smaller the pressure. However,
the power law nature of this relation is not trivial. For the simulations, we also find an excellent
collapse of $P$ vs. $1-f_{NR}$, via a power-law, although the dynamics deviates from power law scaling at large nonrattler fractions, and note a small deviation in the overall pressure scale. A quantitative magnitude agreement in particular in the pressure is sensitive to the choice of inclusion of rattlers, the force law used in the numerics and the area normalization choice in the Irving-Kirkwood formalism, and hence it is more challenging to get consistent among the experimental and numerical data sets. However, the trends of $f_{NR}$ collapse and $P(f_{NR})$ are mostly insensitive to these differences between experiments and simulations.

Figure~\ref{fig:23}b shows the stress anisotropy, given by $\tau/P$.   We find an initial rapid increase of $\tau$ from the randomly prepared nominally isotropic stress-free initial state.  After this transient, the anisotropy shows only a modest decrease with $f_{NR}$ for both experiments and simulations, while it remains nonzero.  The decreasing trend is consistent with the observation that shear jammed states initially have a very anisotropic network, that evolves towards a more isotropic one with increasing strain. The agreement between experiments and  simulations is only qualitative, but even though there is modest scatter in the numerical data, the collapse with $f_{NR}$ is obvious.

\emph{Force Probability Distribution ---} A useful microscopic measure is the probability distribution function (PDF) of the norm of the contact forces $P(|\textbf{f}|)$. Much work has been devoted to characterizing and understanding this distribution~\cite{corwin2005, tighe2011}  although isotropically compressed packings have been the primary focus of past work.  Here we examine $P(|\textbf{f}|)$ vs. $f_{NR}$, and contrast the experimental and numerical data.
Figure~\ref{fig:23}c shows the probability of finding a contact force of a certain magnitude in the entire data set at $\phi = 0.7863$. We collect the data for all strain values at this density to obtain better statistics. The numerical and experimental histograms are very similar. Also the probability distribution function again shows a good collapse with $f_{NR}$: in Figure~\ref{fig:23}d we show the bin with force centered on $1.0\langle |\textbf{f}| \rangle$.  In both experimental and numerical data, there are good collapses with $f_{NR}$; the experimental and numerical data are in qualitative agreement. {So, this seems to say that the probability of finding $f$ at the mean grows significantly with $f_{NR}$. Does that sound right?} The collapse with $f_{NR}$ is observed for other force bins; for example, we have verified that the same feature are observed for force bin centered on $0.5\langle |\textbf{f}| \rangle$
and $1.5\langle |\textbf{f}| \rangle$ (see Appendix).

\begin{figure}
\includegraphics[width=8.5cm]{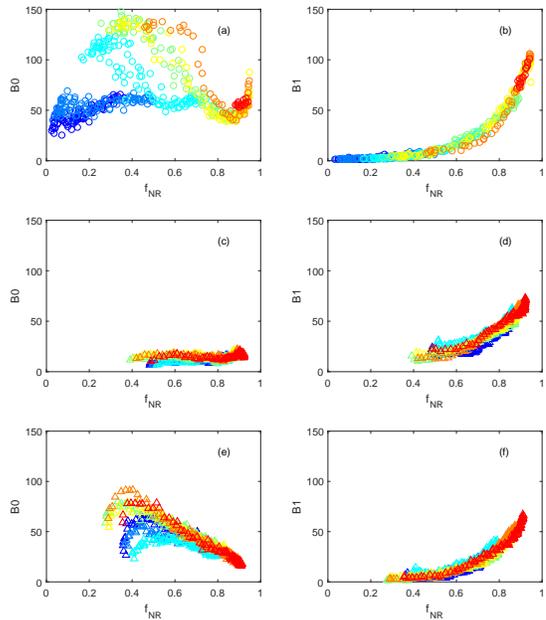}
\caption{\label{fig:4} Betti numbers, $\beta_{0}$ (left) and $\beta_{1}$ (right), as a function of $f_{NR}$ for both experiments (a,b: $\circ$) and simulations (c-f, $\bigtriangleup$). (c,d) show original numerical data; (e,f) show the numerical data with 0.01N noise added.}
\end{figure}

The previous metrics have addressed either macro-scale or micro-scale structural properties, but do not  provide detailed information about the structure of force networks, such as those shown in Fig.~\ref{fig:0}.  We now use topological metrics to analyze the properties of the network that describes the force interactions between the particles with force magnitude larger than $f_c = 1.0\langle |\textbf{f}| \rangle$.

Figure~\ref{fig:4}a,b shows the resulting Betti number dynamics for the experiments. $\beta_0$ shows a plateau around $100$ and subsequently a decrease. Simulations without added noise (Figure~\ref{fig:4}c) however, show very  different results with much smaller values of $\beta_0$: they are essentially independent of $f_{NR}$. This observation is in a sharp contrast to the findings for more conventional measures, for which experimental and numerical simulations showed at least a qualitative agreement. We see a similar difference in the $\beta_1$ dynamics in Figure~\ref{fig:4}b,d: the numerical data do not seem to asymptote to $\beta_1 = 0$ in the limit of $f_{NR} \rightarrow 0$.

The question is: what causes such a dramatic  difference in the numerical and experimental network properties? To explain the larger number of  the connected components present in the experimental data we propose the following mechanism. Consider the part of the contact network at which the particle interactions are a little below $f_c$. A slight increase of a single interaction in this region will introduce a new connected component. On the other hand a slight decrease of single interaction, at the region where the interactions are just a little above $f_c$ might result in  breaking  a connected component into two pieces. Noise will thus affect topological features in the network.
To explore this possibility, we add random noise to the simulations results.   The random noise that we use is chosen from a flat distribution within the range of $[0, 0.01]$N (so with mean of $0.005$N) that is consistent with the expected level of error in the  experiments --- see Fig.~\ref{fig:23}c. Note that after the noise is added to the contact forces, a small force inbalance may result on the particles.   The non-zero mean of the added noise is motivated by the fact  that we consider here a (positive definite) norm of the force vector.

After noise addition, the agreement between experimental and simulations improves significantly for both
$\beta_0$ and $\beta_1$.  Figures~\ref{fig:4}e shows a monotonically decreasing behavior of
$\beta_0$, much more in line with experimental data. For $\beta_1$ we observe {\em quantitative} agreement between experiments and simulations. Again, for both $\beta_0$ and $\beta_1$ we see a good collapse when we consider these quantities as functions of $f_{NR}$.

\section{Conclusions} In this paper, we have presented one  of the first attempts to directly compare microscopic contact force level data from experiments and simulations of dense granular systems. We perform the comparison with metrics across the scales that range from micro to meso and macro.  The comparison of micro and macro measures is mostly satisfactory.  In particular, we find for both experiments and simulations that, when we express  the contact number, pressure,
anisotropy, or probability density function of the normal forces as a function of the
non-rattler fraction, $f_{NR}$, we obtain collapse onto master curves,
capturing the dynamics over a wide range of conditions.

After finding good agreement for the quantities specified above, it is perhaps
surprising to see that  the force network in the experiments and simulations, described
by Betti numbers, are qualitatively different when raw data from simulations are
considered. We argue that contact force noise can be of significant influence on mesoscopic network features, and test this by including adding artificial noise to simulation results. We find that after adding noise of small magnitude, the mesoscopic Betti number dynamics is very similar between experiments and numerics.

On the one hand, the fact that noise influences the properties of
force networks, but not classical micro and macro measures listed above  is
encouraging, since it suggests that the quantities that are commonly observable in
granular experiments should not be influenced by typical (experimental) noise.
On the other hand, the fact that noise had to be added to simulation results to reach
agreement with experiments suggests that comparison of the force network
properties is more subtle.   However, sensitivity of the force networks' properties
to noise opens the door towards the use of topological measures to quantify the experimental
noise level and its properties, and to distinguish intrinsic fluctuations from experimental noise.

Our future work will consider in detail other measures quantifying the force networks such as total persistence and related measures that we have considered in our previous works~\cite{kramar2013,physicaD14,poly2} to provide even better understanding of the properties of force networks and their connection to macroscopic
response of particulate-based systems.

\section{Appendix: Experiments}

We discuss here experimental data obtained in a setup described in detail elsewhere~\cite{ren13, ren11}. We use $\sim 1000$ bi-disperse photoelastic particles in a linear shear cell with articulated bottom. The articulated bottom shear induces a linear shear profile, suppressing shear bands and other inhomogeneities. We extract the local particle stress by either a nonlinear
pattern-fitting algorithm discussed previously~\cite{majmudar07,zhang10,bi11} that yields the complete contact network, particle forces, and stress tensor (e.g. pressure $P$ and anisotropy $\tau$), or via $G^2$, the local squared intensity gradient of the photoelastic response, averaged on each particle. $G^2$ is a one-to-one function of $P$ on the particle level, and provides an easy measure for $P$. For small data sets, we use the former approach; for larger data sets, we use the latter approach to get only $P$. In addition to $P$ and $\tau$, we can probe the contact number dynamics with great accuracy: due to the photoelastic response of the disks, we can determine contacts with much more sensitivity than typical distance based metrics~\cite{arevalo14}. We can thus measure the fraction of particles with $n$ contacts, which we describe with $Z_n$. We thus also have access to the number of non-rattlers $f_{NR}$, the particles that are in a force bearing network. Recent work~\cite{bi11} showed that this was an important microstructural metric. We probe these structural metrics and force networks in shear jammed states by shearing the system. We prepared packings in a stress free initial state, for $0.75 \leq \phi \leq 0.825$. We then quasi-statically shear the system by 100~small strain steps of 0.27\%, up to a total strain of $\gamma = 27\%$
~\cite{ren11}. For the larger $\phi$'s considered here, we could not apply the full $27\%$ strain because $P$ became so large that the layer of disks was unstable to out-of-plane buckling.  If buckling occurs, we terminate the shear experiment. At each packing fraction, we carried out the same shear experiments five times for reproducibility. Beyond a contact pressure of about 10 $N/m$ or, equivalently, $f_{NR} = 0.95$, the compression of the disks is such that the particles visibly deform, and hence the force inversion method breaks down. We measure the quality of force extraction by calculating the total image difference between the original photoelastic images and their reconstructed equivalents, where the reconstructed image is based on the fitted contact forces. We show the results in Fig.~\ref{fig:imdiff}. At large $f_{NR}$ we can clearly see that the image reconstruction becomes significantly different from the image as taken during the experiment, so we omit all data for $f_{NR} > 0.95$.

\begin{figure}
\includegraphics[width=8.5cm]{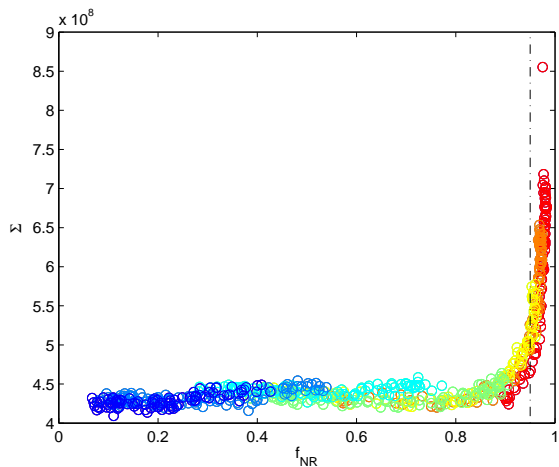}
\caption{\label{fig:imdiff} Integrated image difference $\Sigma$ for the averaged experimental runs at different densities. There is a constant background difference due to pixel level noise and other artifacts. The dashed line at $f_{NR} = 0.95$ represents the cutoff we impose.}
\end{figure}

\section{Appendix: Numerics}

We perform discrete element simulations using a set of circular disks confined in an initially rectangular domain. The number of the particles ranges from $910$ up to $1050$, depending on the packing fraction and is set exactly to the number of particles in the experiments.  The walls are composed of monodisperse disks. The domain is rectangular, and the length and width of the rectangle are $54$ and $27$ particle diameters, respectively. System particles are bidisperse and the ratio of the diameters of the large and small particles is $\approx15.9/12.7$. The exact positions and particle radii for all packing fractions and different realizations are taken from the initial conditions in the experiments. Particles are soft and interact via normal and tangential forces during collision, with static friction and viscous damping.

\begin{figure}
  \includegraphics[width=8.5cm]{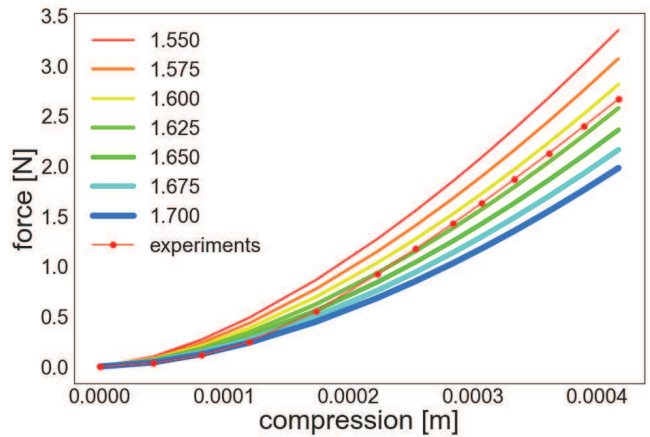}
  \caption{Calibration curve from experiments (o) and curves using non-linear force model in Equation~\ref{eq:non_lin_fm} with different $\delta$. }\label{fig:force_calibration}
\end{figure}

The force model used in the simulations is non-linear; normal force between particles $i$ and $j$ is (see~\cite{kondic_99} for more details) given by
\begin{eqnarray}
 {\bf F}_{i,j}^n& =&k_nx^{\delta} {\bf n} - \gamma_n x^{0.5} \bar m {\bf v}_{i,j}^n \label{eq:non_lin_fm}\\
 r_{i,j} &=& |{\bf r}_{i,j}|, \ \ \ {\bf r}_{i,j} = {\bf r}_i - {\bf r}_j, \ \ \  {\bf n} = {\bf r}_{i,j}/r_{i,j}\nonumber \\
 k_n &=& {2Y\over 3(1-\sigma^2)}d_{ave}^{1-\beta} \nonumber
\end{eqnarray}
where $1+\beta=\delta$, ${\bf v}_{i,j}^n$ is set to the relative normal velocity and $Y$ and $\sigma$ are Young's modulus and Poisson ratio, respectively. The amount of compression is $x = d_{i,j}-r_{i,j}$, where $d_{i,j} = {(d_i + d_j)/2}$; $d_{i}$ and $d_{j}$ are the diameters of the particles $i$ and $j$; $d_{ave}$ is the average particle diameter. Here, ${\bf r}_i, {\bf r}_j$
are the vectors pointing from the centers of particles $i,j$ towards the point of contact.

The exponent $\delta$ in the force model is chosen to match the experimental calibration curve measuring the amount of compression of the photoelastic disk particle pressed by a steel plate with a given force. Figure~\ref{fig:force_calibration} shows the experimental calibration curve and $k_nx^{\delta}$ curves for chosen values of $\delta$ with $k_n$ (dependent on $\delta$, as specified in the set of Equations~\ref{eq:non_lin_fm}) and $Y=3.45~\rm{GPa}$ and $\sigma=0.5$ set to match the experiments. We find that choosing $\delta=1.625$ yields the least square difference between $k_nx^{\delta}$
and experimental calibration curve and therefore we use this value of $\delta$ in the force model for particle-particle interaction in the simulations.

For the simulations, the characteristic length scale is {\bf $d_{ave}$}, the average particle mass, $\bar m$,
is the mass scale and the binary particle collision time  $\tau_c$
is the time scale. The value of $\tau_c$ is set to~\cite{kondic_99}
\begin{equation}
  \tau_c = \alpha (1+0.5\beta)^{1\over 2+\beta} \Biggl(\bar m {3(1-\sigma^2)\over 2Yd_{ave}^{(1-\beta)}}\Biggr)^{1\over 2+\beta}v_0^{-\beta\over2+\beta}
\end{equation}
where $v_0=0.01423 ~\rm{ms^{-1}}$ is a characteristic magnitude of velocity in the system (shearing speed); prefactor $\alpha$ and damping coefficient $\gamma_n$ is obtained using the coefficient of restitution, $e=0.5$, as reported in~\cite{kondic_99}.

\vskip -0.02in

We implement the commonly used Cundall-Strack model for static friction~\cite{DEM_ori}, where a tangential spring is introduced between particles for each new contact that forms at time $t=t_0$. Due to the relative motion of the particles, the spring length, ${\boldsymbol\xi}$, evolves as $\boldsymbol\xi=\int_{t_0}^t {\bf v}_{i,j}^t~(t')~dt'$, where ${\bf v}_{i,j}^{t}= {\bf v}_{i,j} - {\bf v}_{i,j}^n$.  For long lasting contacts, $\boldsymbol\xi$ may not remain parallel to the current tangential direction defined by $\bf{t}={\bf v}_{i,j}^t/|{\bf v}_{i,j}^t|$
(see, e.g,.~\cite{brendel98}); we therefore define the corrected $\boldsymbol\xi{^\prime} = \boldsymbol\xi - \bf{n}(\bf{n} \cdot \boldsymbol\xi)$ and introduce the test force
\begin{equation}
{\bf F}^{t*} =-k_tx^{\beta}\boldsymbol\xi^\prime - \gamma_t x^{0.5} \bar m {\bf v}_{i,j}^t
\end{equation}
where $k_t = 6/7k_n$ (close to the value used in~\cite{goldhirsch_nature05}), $\gamma_t$ is the coefficient of viscous damping in the tangential direction (with $\gamma_t = {\gamma_n}$). The value of $\mu = 0.7$ is set to the inter-particle friction from the experiments~\cite{renthesis}. To ensure that the magnitude of the tangential force remains below the Coulomb threshold, we constrain the tangential force to be
\begin{equation}
 {\bf F}^t = min(\mu |{\bf F}^n|,|{\bf F}^{t*}|){{\bf F}^{t*}/|{\bf F}^{t*}|}
\end{equation}
and redefine ${\boldsymbol\xi}$ if appropriate. In the force model, we include interaction with the base. The force between the particle and the base has a translational and a  rotational component and the particle-base friction coefficient is $\mu_b=0.4$, corresponding to the reported experimental value~\cite{renthesis}. The magnitude of the deceleration of the particle in translational direction due to the friction with base is $\mu_b|{\bf g}|$ where ${\bf g}$ is the gravitational acceleration. The rotational deceleration of $i$-th particle due to friction with base
\begin{equation}
 |\alpha_i|={4\over3}\mu_b{|{\bf g}|\over r_i}
\end{equation}
is computed by integrating torque arising from friction with the base and using the moment of inertia of the disk, $I=(m_ir_i^2)/2$, where $m_i,r_i$ are the values of mass and radius of the $i$-th particle, consecutively. For simplicity, we use $r_i=1/3d_{ave}$ for both small and large particles.
We integrate Newton's equations of motion for both the translation and rotational degrees of freedom using a $4$th order predictor-corrector method with time step $\Delta t =0.02\tau_c$. From the initial configuration taken from the experiments, the system is sheared by moving the left wall in positive and the right wall in the negative direction. Shearing speed used in the simulations, expressed in the units of $d_{ave}/\tau_c$, is $v_0'=2.5\times 10^{-5}$. Relaxation is interjected after each strain step of $0.27\%$. The maximum strain amplitude is $27\%$.

\begin{figure}[ht!]
\includegraphics[width=8.5cm, viewport= 0 200 550 500]{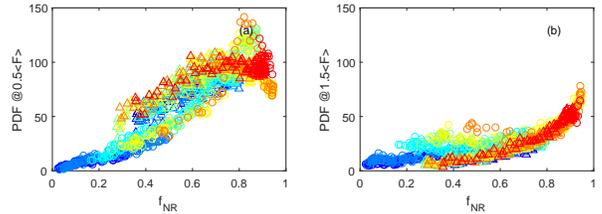}
\caption{Probability distribution function of contact forces for a force magnitude bin centered at $0.5\langle|\textbf{f}| \rangle$ (left) and $1.5\langle|\textbf{f}| \rangle$ (right)}\label{fig:imdiff}
\end{figure}
\begin{figure}[ht!]
\includegraphics[width=8.5cm]{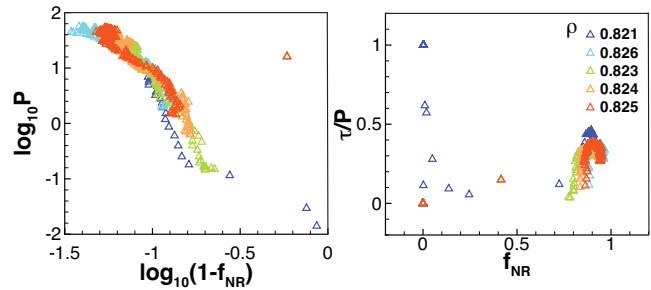}
\caption{Pressure, $\log_{10}(P)$ and anisotropy, $\tau/P$ as a function of $\log_{10}(1-f_{nr})$ and $f_{nr}$, respectively.}\label{fig:frictionless}
\end{figure}

\section{Appendix: Consistency checks}

In this section we present consistency checks; specifically, we focus on the distribution of the forces measured in both experiments and simulations for a different force bin than in the main body of the paper. Then we show that the particle-particle and particle-base friction is essential if we want to achieve a quantitative agreement between simulations and experiments.

In Figure~3d we show the probability distribution function of contact forces for a force magnitude bin of $1.0\langle|\textbf{f}| \rangle$. In Figure~\ref{fig:imdiff} we show that the $f_{NR}$ collapse and consistency between experimental and numerical data is retained for the force bin around $0.5\langle|\textbf{f}| \rangle$ (a) and $1.5\langle|\textbf{f}| \rangle$ (b). Thus we conclude that our results comparing force distribution is not sensitive to the choice of the bin.

To demonstrate the importance of having a non-zero friction coefficient, we
 consider the dynamics of pressure and anisotropy for a few single runs with $\mu=0.0$ and $\mu_b=0.0$ at various different packing fractions around the jamming point as observed in our protocol. Figure~\ref{fig:frictionless} shows $P$ and $\tau/P$ as a function of $f_{NR}$ for frictionless systems and should be compared to Fig.~3a,b
 Evidently, our numerical protocol does not produce the same $f_{NR}, P$ dynamics as the experiments, even though the pressure and anisotropy values remain within a similar range in jammed systems (especially for large $f_{NR}$).

\emph{Acknowledgments}
*Dijksman and Kovalcinova contributed equally to this work. This work was partially supported by
NSF-DMS-0835621, 0915019, 1125174, AFOSR Grant Nos. FA9550-09-1-0148,
FA9550-10-1-0436 and DARPA (M.K., and K.M) HR0011-16-2-0033 (K.M, L.K, R.P.B) and NSF Grant
No. DMS-0835611, and DTRA Grant No. 1-10-1-0021; NSF Grants
DMR-12063251 and DMS1248071 and NASA grants NNX10AU01G and
NNX15AD38G; DARPA grant HR0011-16-2-0033 (KM)

\end{document}